\documentclass[twoside]{ilcws10}
\usepackage[latin1]{inputenc}
\usepackage[dvips]{graphicx,epsfig,color}
\usepackage{wrapfig,rotating}
\usepackage{amssymb,amsmath,array}

\begin{document}
\title{
Present Status of GRACE/SUSY-loop} 
\author{M. Jimbo$^1$,  K. Iizuka$^2$, T. Ishikawa$^3$, K. Kato$^4$,
T. Kon$^2$, Y. Kurihara$^3$ and M. Kuroda$^5$
\vspace{.3cm}\\
1 Chiba University of Commerce, Ichikawa, Chiba 272-8512, Japan
\vspace{.1cm}\\
2 Seikei University, Musashino, Tokyo 180-8633, Japan
\vspace{.1cm}\\
3 KEK, Tsukuba, Ibaraki 305-0801, Japan
\vspace{.1cm}\\
4 Kogakuin University, Shinjuku, Tokyo 163-8677, Japan
\vspace{.1cm}\\
5 Meiji Gakuin University, Yokohama, Kanagawa 244-8539, Japan
}

\maketitle

\begin{abstract}
{\tt GRACE/SUSY-loop} is a program package for the automatic calculation of
the MSSM amplitudes in one-loop order.  We present features of {\tt GRACE/SUSY-loop},
processes calculated using {\tt GRACE/SUSY-loop} and an extension of
the non-linear gauge formalism applied to {\tt GRACE/SUSY-loop}.
\end{abstract}

\section{Introduction}

Despite its compactness and success in describing known experimental data available
up to now, the standard model (SM) is considered to be an effective theory valid only
at the presently accessible energies on account of theoretical problems.  Supersymmetric
(SUSY) theory, which predicts the existence of a partner to every particle of the SM that
differs in spin by one half, is believed to be an attractive candidate for the theory
beyond the SM (BSM).  The minimal supersymmetric extension of the SM (MSSM) remains
consistent with all known high-precision experiments at a level comparable to the SM.
One of the most important aim of the particle experiments at sub-TeV-region and
TeV-region energies is to probe evidence of the BSM, so search for SUSY particles
plays crucial role in it.

Experiments at present and future accelerators, the Large Hadron Collider (LHC) and
the International Linear Collider (ILC), are expected to discover SUSY particles and
provide accurate data on them.  In particular, experiments at the ILC offer high-precision
determination of SUSY parameters via $e^- e^+$-annihilation processes.  Since the
theoretical predictions with the similarly high accuracy is required for us to extract
important physical results from the data, we have to include at least one-loop
contributions in perturbative calculations of amplitudes.

Among SUSY particles, only the lightest one (LSP) is stable if $R$-parity between usual
particles and their SUSY partners is conserved, and the others decay without exception.
Then decay processes should be analyzed precisely in experiments at the LHC and the ILC.
Recently, we have calculated the radiative corrections to production processes and decay
processes of SUSY particles in the framework of the MSSM using
{\tt GRACE/SUSY-loop}~\cite{Fujimoto:2007bn, Jimbo:2008, Iizuka:2008, Iizuka:2010bh, Kon:2010}.
In this paper, we show features of {\tt GRACE/SUSY-loop}, processes calculated using
{\tt GRACE/SUSY-loop} and an extension of the non-linear gauge (NLG)
formalism~\cite{Fujikawa:1973, Gavela:1981, Haber:1988, Capdequi:1990, Boudjema:1996, Kato:2006}
applied to {\tt GRACE/SUSY-loop}~\cite{Fujimoto:2006km}.

\section{Features of GRACE/SUSY-loop}

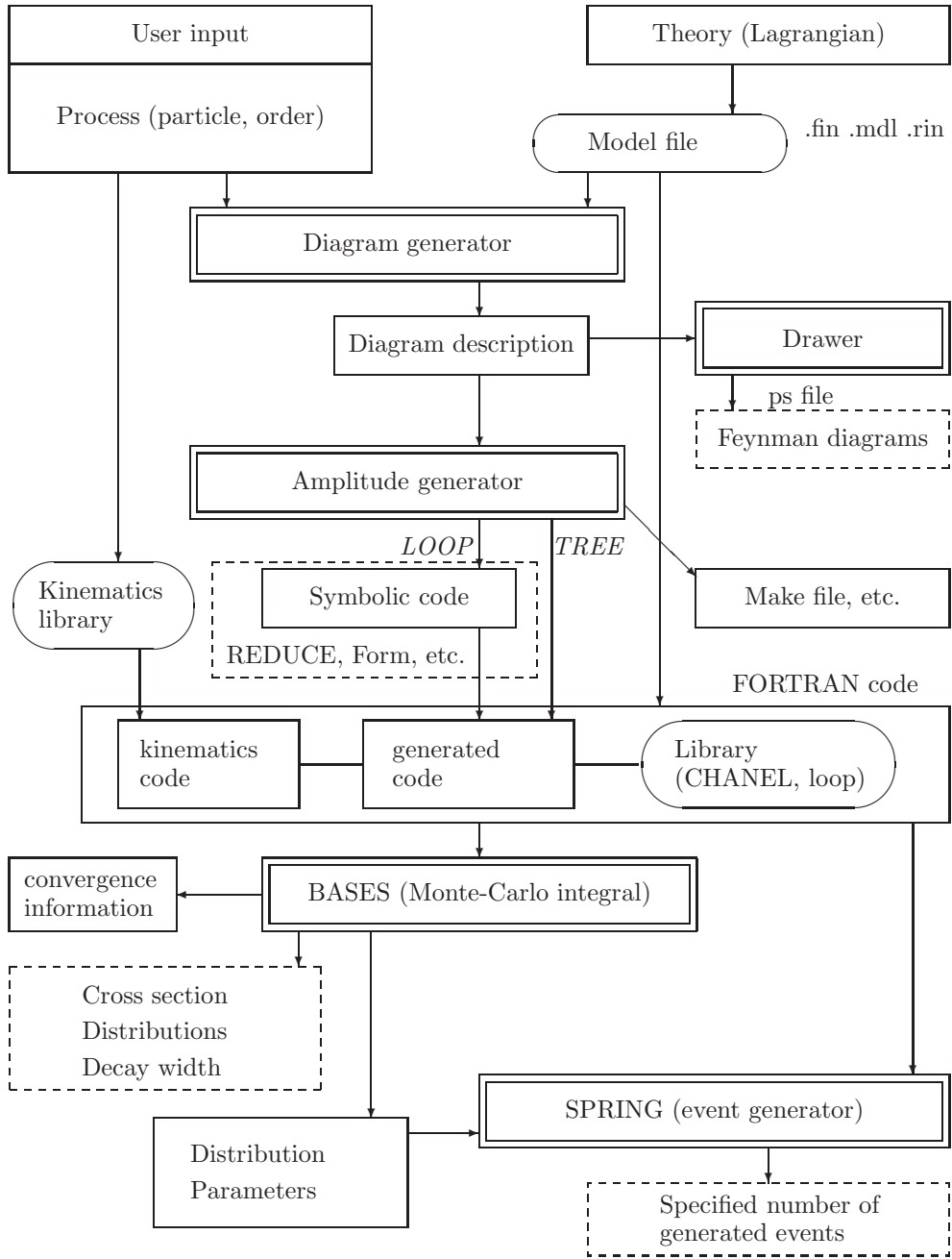
\begin{figure}[htbp]
\setlength{\unitlength}{1mm}
\begin{picture}(120,180)

\put(85,170){\framebox(50,8)[c]{\hbox{Theory (Lagrangian)}}}
\put(5,170){\framebox(50,8)[c]{\hbox{User input}}}
\put(5,155){\framebox(50,15)[c]{\hbox{Process (particle, order)}}}

\put(95,159){\oval(35,8)[c]{}}
 \put(85,158){\hbox{Model file}}

\put(115,160){\makebox{.fin .mdl .rin}}

\put(105,170){\vector(0,-1){7}}
\put(95,155){\vector(0,-1){74}}
\put(85,155){\vector(0,-1){5}}
\put(35,155){\vector(0,-1){5}}
\put(20,155){\vector(0,-1){54}}

\put(31,141){\framebox(58,8)[c]{\  }}
\put(30,140){\framebox(60,10)[c]{\hbox{Diagram generator}}}

\put(70,140){\vector(0,-1){5}}
\put(70,127){\vector(0,-1){10}}

  \put(50,127){\framebox(35,8)[c]{\hbox{Diagram description}}}
  \put(101,128){\framebox(33,8)[c]{ }}
  \put(100,127){\framebox(35,10)[c]{\hbox{Drawer}}}
  \put(100,114){\dashbox(35,8)[c]{\hbox{Feynman diagrams}}}
    \put(85,132){\vector(1,0){15}}
    \put(105,127){\vector(0,-1){5}}

\put(110,123){\makebox{ps file}}

  \put(18,95){\oval(25,12)[c]{}}
   \put(9,96){\hbox{Kinematics}}
   \put(9,92){\hbox{library}}
  \put(23,89){\vector(0,-1){10}}

\put(31,108){\framebox(58,8)[c]{\  }}
\put(30,107){\framebox(60,10)[c]{\hbox{Amplitude generator}}}
\put(100,92){\framebox(35,8)[c]{\hbox{Make file, etc.}}}
  \put(90,109){\vector(1,-1){10}}
  \put(80,107){\vector(0,-1){28}}

\put(59,102){\makebox{\it LOOP}}
\put(80,102){\makebox{\it TREE}}

\put(33,85){\dashbox(45,16){\  }}
\put(40,92){\framebox(35,8)[c]{\hbox{Symbolic code}}}
  \put(70,107){\vector(0,-1){7}}
  \put(70,92){\vector(0,-1){13}}

\put(35,87){\makebox{REDUCE, Form, etc.}}

\put(105,83){\makebox{FORTRAN code}}

  \put(15,65){\framebox(120,16)[c]{\  }}
    \put(20,67){\framebox(25,12)[c]{\  }}
    \put(54,67){\framebox(29,12)[c]{\  }}
    \put(23,70){\hbox{code}}
    \put(23,74){\hbox{kinematics}}
    \put(58,70){\hbox{code}}
    \put(58,74){\hbox{generated}}
    \put(45,73){\line(1,0){9}}
    \put(92,73){\line(-1,0){9}}

  \put(110,73){\oval(35,12)[c]{}}
    \put(97,70){\hbox{(CHANEL, loop)}}
    \put(97,74){\hbox{Library}}
    
\put(70,65){\vector(0,-1){5}}

\put(41,51){\framebox(58,8)[c]{\  }}
\put(40,50){\framebox(60,10)[c]{\hbox{BASES (Monte-Carlo integral)}}}

  \put(5,50){\framebox(23,10){\  }}
    \put(7,52){\hbox{information}}
    \put(7,56){\hbox{convergence}}
  \put(40,55){\vector(-1,0){12}}

  \put(5,28){\dashbox(43,17){\  }}
    \put(15,30){\hbox{Decay width}}
    \put(15,35){\hbox{Distributions}}
    \put(15,40){\hbox{Cross section}}
  \put(45,50){\vector(0,-1){5}}

  \put(55,50){\vector(0,-1){26}}
  \put(25,9){\framebox(35,15){\  }}
    \put(30,13){\hbox{Parameters}}
    \put(30,18){\hbox{Distribution}}
  \put(60,22){\vector(1,0){10}}

\put(71,21){\framebox(63,8)[c]{\  }}
\put(70,20){\framebox(65,10)[c]{\hbox{SPRING (event generator)}}}

\put(130,65){\vector(0,-1){35}}
\put(110,20){\vector(0,-1){5}}

  \put(85,5){\dashbox(50,10){\  }}
    \put(95,7){\hbox{generated events}}
    \put(95,11){\hbox{Specified number of}}
\end{picture}
\caption{The system flow of {\tt GRACE}}\label{Fig:Flow}
\end{figure}

For many-body final states, each production process or decay process is described by
a large number of Feynman diagrams even in tree-level order.  There are still more
Feynman diagrams in one-loop order even for two-body final states.  For this reason,
we have developed the {\tt GRACE} system~\cite{Yuasa:1999rg}, which enables us to calculate
amplitudes automatically.  Figure \ref{Fig:Flow} shows the system flow of {\tt GRACE}
generically.  A program package called {\tt GRACE/SUSY-loop} is the version of the {\tt GRACE}
system for the calculation of the MSSM amplitudes in one-loop order, which includes
the model files of the MSSM and the loop library.  There exist other program packages
developed by other groups independently for the calculation of the MSSM amplitudes
in one-loop order, {\tt SloopS}~\cite{Baro:arXiv0906.1665} and
{\tt FeynArt/Calc}~\cite{Hahn:2000jm}.

\subsection{Renormalization scheme}

As explained in \cite{Fujimoto:2007bn}, the renormalization scheme adopted for
the electroweak (EW) sector in {\tt GRACE/SUSY-loop} is a variation of the on-mass-shell
scheme~\cite{Chankowski:1994, Yamada:1991, Dabelstein:1995, Hollik:2002, Fritzsche:2002},
which is a MSSM extension of the scheme in the SM used in {\tt GRACE-loop}~\cite{Yuasa:1999rg}.  
We impose the on-mass-shell condition on gauge bosons ($W$, $Z$, $\gamma$), all fermions ($f$),
sfermions ($\widetilde{f}$), CP odd Higgs ($A^0$), the heavier CP even Higgs ($H^0$), both charginos
($\widetilde\chi_1^+$, $\widetilde\chi_2^+$), and the lightest neutralino ($\widetilde\chi_1^0$),
then these particles have no mass correction in one-loop order of the EW sector.  There are some
freedom in the renormalization scheme of the sfermion sector.  They are distinguished by different
choice of residue conditions, decoupling conditions on the transition terms between the
lighter and the heavier sfermions, and left-handed SU(2) relations in one-loop order.
Recently, our calculations have been performed with the scheme in which the residue conditions
are imposed on all sfermions except for the heavier stop and sbottom
($\widetilde{t}_2$, $\widetilde{b}_2$).  Corrections of the external line for
$\widetilde{t}_2$ and $\widetilde{b}_2$ become non-zero in this scheme.

The renormalization schemes adotpted for the QCD sector are separate ways between light
and massive particles.  Light quarks in the first and second generation and gluon are treated in the
$\overline{DR}$ scheme~\cite{Siegel:1979, Capper:1979} as in the convensional perturbative QCD.
Massive quarks in the third generation and gluino are handled by the on-mass-shell scheme as in
the EW sector.  For the regularization of infrared divergences, the fictitious mass of gluon
$\lambda$ is used in the previous version of {\tt GRACE/SUSY-loop}~\cite{Fujimoto:2007bn}.
We have developed a new version of the system in which mass-singularities are regularized by
the dimensional method~\cite{Iizuka:2010bh}.  In order to refer the ultraviolet and the infrared
divergences, we define the notations $C_{\rm UV} \equiv 1/\epsilon$ and
$C_{\rm IR} \equiv 1/\,\overline\epsilon$, where the dimension of the space-time $d$ is related to
$\epsilon$ and $\overline\epsilon$ as $d = 4 - 2\epsilon = 4 + 2\,\overline\epsilon$.

\subsection{Non-linear gauge formalism}

In {\tt GRACE/SUSY-loop}, we use the technique of the NLG formalism in order to
confirm the validity of calculations by imposing the NLG invariance on physical results.
The NLG formalism is an extension of the linear $R_\xi$-gauge.  The gauge fixing lagrangian for
the EW interactions in the linear $R_\xi$-gauge is as follows:
\begin{eqnarray}
  {\cal L}_{\rm gf} &=&  -{1\over{\xi_W}}\vert F_{W^+} \vert^2 
                      -{1\over{2\xi_Z}}(F_Z)^2
                      -{1\over{2\xi_\gamma}}(F_\gamma)^2, \label{L1}\\
   F_{W^\pm} &=& \partial_\mu W^{\pm\mu} \pm i\xi_W M_WG^\pm ,\label{L2}\\
   F_Z &=& \partial_\mu Z^\mu+\xi_ZM_ZG^0, \label{L3} \\
   F_\gamma &=& \partial_\mu A^\mu, \label{L4}
\end{eqnarray}
where $G^\pm$ and $G^0$ stand for the Goldstone bosons which correspond to
gauge bosons $W^\pm$ and $Z$, respectively.

The following NLG functions are introduced to the EW sector of the MSSM Lagrangian,
\begin{eqnarray}
  F_{W^\pm} &=& (\partial_\mu \pm ie\tilde\alpha A_\mu  
                  \pm ig c_W\tilde\beta Z_\mu) W^{\pm\mu}
            \pm i\xi_W {g\over 2}(v + \tilde\delta_H H^0 + \tilde\delta_h h^0 
              \pm i\tilde\kappa G^0)G^\pm , \label{N1}\\
   F_Z &=& \partial_\mu Z^\mu+\xi_Z{{g_Z}\over 2}(v +\tilde\epsilon_H H^0
                +\tilde\epsilon_h h^0)G^0, \label{N2} \\
   F_\gamma &=& \partial_\mu A^\mu, \label{N3}
\end{eqnarray}
where $v = \sqrt{v_1^2+v_2^2}$ , $M_W = g v / 2$ , $M_Z = g_Z v / 2$ ,
and $h^0$ stands for the lighter CP even Higgs.
They contain seven independent NLG-parameters,
($\tilde\alpha, \tilde\beta, \tilde\delta_H, \tilde\delta_h, \tilde\kappa, 
\tilde\epsilon_H, \tilde\epsilon_h$).
We perform the numerical tests by varying these parameters.

\subsection{Tests of numerical results}

Since the {\tt GRACE} system provides numerical results automatically, we have to
test validity of the results.  We adopt four tests for the EW sector and three tests
for the QCD sector as in Table \ref{tab:tests}.
\begin{table}[htb]
\centerline{\begin{tabular}{lll}
\hline
& Tests  & Variables \\\hline
\rule{0mm}{11.5pt}
EW: & NLG invariance  & $\tilde\alpha, \tilde\beta, \tilde\delta_H, \tilde\delta_h,
\tilde\kappa, \tilde\epsilon_H, \tilde\epsilon_h$ \\
& Cancellation of ultraviolet divergence & $C_{\rm UV}$ \\
& Cancellation of infrared divergence & $\lambda$ (fictitious mass of photon) \\
& Independence of the cutoff energy of the soft photon & $k_{\rm c}$ \\
\hline
\rule{0mm}{11pt}
QCD: & Cancellation of ultraviolet divergence & $C_{\rm UV}$ \\
& Cancellation of infrared divergence & $C_{\rm IR}$ \\
& Independence of the cutoff energy of the soft gluon & $k_{\rm c}$ \\
\hline
\end{tabular}}
\caption{List of tests}
\label{tab:tests}
\end{table}

\section{Calculated processes}

We have calculated the radiative corrections to production processes and decay processes
of SUSY particles in the framework of the MSSM using {\tt GRACE/SUSY-loop}.  Table
\ref{tab:proc} shows the list of processes calculated using {\tt GRACE/SUSY-loop}.
In \cite{Iizuka:2008}, we have calculated the radiative corrections of sfermion-decay
processes using the parameter set adopted in \cite{Guasch:2002ez}.
\begin{table}[htb]
\centerline{\begin{tabular}{lcc}
\hline
Processes  & {\tt GRACE} & Preceding studies \\\hline
Chargino-pair production ($e^- + e^+ \rightarrow \widetilde{\chi}^-_1 + \widetilde{\chi}^+_1$)
& \cite{Fujimoto:2007bn} & \cite{Fritzsche:2004nf, Oller:2005xg, kilian:2006} \\
Chargino decay ($\widetilde{\chi}^+_2 \rightarrow$ two body and $\widetilde{\chi}^+_1 \rightarrow$
three body) & \cite{Fujimoto:2007bn} & \\
Neutralino-pair production ($e^- + e^+ \rightarrow \widetilde{\chi}^0_1 + \widetilde{\chi}^0_2$)
& \cite{Jimbo:2008} & \cite{Fritzsche:2004nf, Oller:2005xg} \\
Neutralino decay ($\widetilde{\chi}^0_{2,3,4}$ $\rightarrow$ two body and $\widetilde{\chi}^0_2 \rightarrow$
three body) & \cite{Jimbo:2008} & \cite{Drees:2006um} \\
Sfermion decay ($\widetilde{f} \rightarrow$ two body)
& \cite{Iizuka:2008} & \cite{Beenakker:1996dw, Beenakker:1997, Guasch:2002ez} \\
Stop decay ($\widetilde{t}_1 \rightarrow b + \widetilde{\chi}_1^+$, $t + \widetilde{\chi}^0_1$ and
$\widetilde{t}_1 \rightarrow b + W^+ + \widetilde{\chi}^0_1$)
& \cite{Iizuka:2010bh, Kon:2010} & \cite{Beenakker:1996dw, Beenakker:1997, Guasch:2002ez} \\
Gluino decay ($\widetilde{g} \rightarrow b + \widetilde{b}_1$, $t + \widetilde{t}_1$)
& \cite{Iizuka:2010bh} & \cite{Beenakker:1996dw} \\
\hline
\end{tabular}}
\caption{List of processes calculated using {\tt GRACE/SUSY-loop}}
\label{tab:proc}
\end{table}
In \cite{Iizuka:2010bh}, however, we use the SPS1a' parameter set \cite{AguilarSaavedra:2005pw}
for the two-body decay processes of the lighter stop ($\widetilde{t}_1$) and an original parameter
set for the three-body decay process of $\widetilde{t}_1$.

\section{Extension of non-linear gauge formalism}

We can extend the NLG functions in the MSSM (\ref{N1}) and (\ref{N2}) by including bilinear
forms of sfermions with new NLG parameters $\tilde{c}$'s as follows:
\begin{eqnarray}
 F_{W^+} &=& (\partial_\mu + ie\tilde\alpha A_\mu  
                  + ig c_W\tilde\beta Z_\mu) W^{+\mu}
            + i\xi_W {g\over 2}(v + \tilde\delta_H H^0 + \tilde\delta_h h^0 
              + i\tilde\kappa G^0)G^+ \nonumber\\
  &+& i\xi_W g\Bigl[
          \sum_{ij} \left\{ 
           \tilde c^{du}_{ij} (\tilde d_i^* \tilde u_j)
          +\tilde c^{sc}_{ij} (\tilde s_i^* \tilde c_j)
          +\tilde c^{bt}_{ij} (\tilde b_i^* \tilde t_j) \right\} \nonumber \\
  &&      + \sum_i \left\{
           \tilde c^e_i(\tilde e_i^* \tilde \nu_e)
          +\tilde c^\mu_i(\tilde \mu_i^* \tilde \nu_\mu)
          +\tilde c^\tau_i(\tilde \tau_i^* \tilde \nu_\tau) \right\} \Bigr], \label{S1} \\
 F_{W^-} &=& (\partial_\mu - ie\tilde\alpha A_\mu  
                  - ig c_W\tilde\beta Z_\mu) W^{-\mu}
            - i\xi_W {g\over 2}(v + \tilde\delta_H H^0 + \tilde\delta_h h^0 
              - i\tilde\kappa G^0)G^- \nonumber \\
  &-& i\xi_W g\Bigl[
          \sum_{ij} \left\{
           \tilde c^{ud}_{ij} (\tilde u_i^* \tilde d_j)
          +\tilde c^{cs}_{ij} (\tilde c_i^* \tilde s_j)
          +\tilde c^{tb}_{ij} (\tilde t_i^* \tilde b_j) \right\} \nonumber \\
  &&      + \sum_i \left\{
           \tilde c^{e*}_i(\tilde \nu_e^* \tilde e_i)
          +\tilde c^{\mu*}_i(\tilde \nu_\mu^* \tilde \mu_i)
          +\tilde c^{\tau*}_i(\tilde \nu_\tau^* \tilde \tau_i) \right\} \Bigr], \label{S2} \\
 F_Z &=& \partial_\mu Z^\mu+\xi_Z{{g_Z}\over 2}(v +\tilde\epsilon_H H^0
            +\tilde\epsilon_h h^0)G^0 \nonumber \\
  &+& \xi_Zg_Z \Bigl[
          \sum_{ij} \left\{
           \tilde c^{uu}_{ij}(\tilde u_i^*\tilde u_j)
          +\tilde c^{dd}_{ij}(\tilde d_i^*\tilde d_j)
          +\tilde c^{cc}_{ij}(\tilde c_i^*\tilde c_j)
          +\tilde c^{ss}_{ij}(\tilde s_i^*\tilde s_j)
          +\tilde c^{tt}_{ij}(\tilde t_i^*\tilde t_j)
          +\tilde c^{bb}_{ij}(\tilde b_i^*\tilde b_j) \right\} \nonumber \\
  &&      + \tilde c^{\nu_e\nu_e}(\tilde\nu_e^*\tilde \nu_e)
          +\tilde c^{\nu_\mu\nu_\mu}(\tilde\nu_\mu^*\tilde \nu_\mu)
          +\tilde c^{\nu_\tau\nu_\tau}(\tilde\nu_\tau^*\tilde \nu_\tau) \nonumber \\
  &&      + \sum_{ij} \left\{
           \tilde c^{ee}_{ij}(\tilde e_i^*\tilde e_j)
          +\tilde c^{\mu\mu}_{ij}(\tilde \mu_i^*\tilde \mu_j)
          +\tilde c^{\tau\tau}_{ij}(\tilde \tau_i^*\tilde \tau_j) \right\} \Bigr], \label{S3}
\end{eqnarray}
where $\tilde c^{ud}_{ij}=\tilde c^{du*}_{ji}~~ (i,j=1,2)$.

\begin{wraptable}[16]{l}{0.5\columnwidth}
\centerline{\begin{tabular}{l}
\hline
Processes~~~ $(i,j,k,\ell=1,2)$ \\\hline
$\widetilde{\nu}_\tau + \widetilde{\nu}_\tau^* \rightarrow \widetilde{\nu}_\tau + \widetilde{\nu}_\tau^*$ \\
$\widetilde{\nu}_\tau + \widetilde{\nu}_\tau^* \rightarrow \widetilde{\tau}_i + \widetilde{\tau}_j^*$ \\
$\widetilde{\tau}_i + \widetilde{\tau}_j^* \rightarrow \widetilde{\tau}_k + \widetilde{\tau}_\ell^*$ \\
$\widetilde{\nu}_\tau + \widetilde{\nu}_\tau^* \rightarrow \widetilde{\nu}_\mu + \widetilde{\nu}_\mu^*$ \\
$\widetilde{\nu}_\tau + \widetilde{\nu}_\tau^* \rightarrow \widetilde{\mu}_i + \widetilde{\mu}_j^*$ \\
$\widetilde{\nu}_\tau^* + \widetilde{\tau} \rightarrow \widetilde{\nu}_\mu^* + \widetilde{\mu}_j$ \\
$\widetilde{\nu}_\tau + \widetilde{\tau}^* \rightarrow \widetilde{\nu}_\mu + \widetilde{\mu}_j^*$ \\
$\widetilde{\tau}_i + \widetilde{\tau}_j^* \rightarrow \widetilde{\mu}_k + \widetilde{\mu}_\ell^*$ \\
$e^- + e^+ \rightarrow \widetilde{\tau}^-_i + \widetilde{\tau}^+_j$ \\
$e^- + e^+ \rightarrow \tau^- + \widetilde{\tau}^+_1 + \widetilde{\chi}^0_1$ \\
$e^- + e^+ \rightarrow \tau^- + \tau^+ + \widetilde{\chi}^0_1 + \widetilde{\chi}^0_1$ \\
\hline
\end{tabular}}
\caption{List of tested processes}
\label{tab:NLG}
\end{wraptable}

We have calculated the cross sections of the processes listed in Table \ref{tab:NLG} in tree-level order,
and confirmed the NLG invariance of the results on the parameters $\tilde{c}$'s in (\ref{S1}), (\ref{S2})
and (\ref{S3}).  Then we are convinced that the extended NLG formalism is valid as a tool to test
the numerical calculations.  

\begin{wrapfigure}[14]{r}{0.5\columnwidth}
\centerline{\includegraphics[width=0.45\columnwidth]{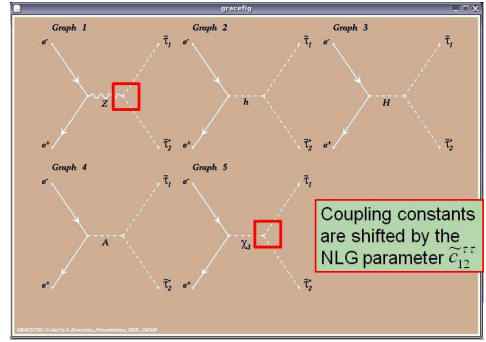}}
\caption{Feynman diagrams of the process
$e^- + e^+ \rightarrow \widetilde{\tau}^-_1 + \widetilde{\tau}^+_2$
in tree-level order.}
\label{Fig:Fd}
\end{wrapfigure}

Figure \ref{Fig:Fd} shows Feynman diagrams of the process $e^- + e^+ \rightarrow \widetilde{\tau}^-_1 + \widetilde{\tau}^+_2$
in tree-level order (drawn by {\tt gracefig}), as an example.  For this process, coupling constants of the vertices,
$\widetilde\tau_1 - \widetilde\tau_2 - Z$ and $\widetilde\tau_1 - \widetilde\tau_2 - G^0 (\chi_3)$ are
shifted by varying the NLG parameter $\tilde c_{12}^{\tau\tau}$, but the sum of the amplitudes is
invariant. Here we set parameters as $M_{\widetilde\tau_1}$ = 320 GeV, $M_{\widetilde\tau_2}$ = 370 GeV,
and total energy as $\sqrt s$ = 1000 GeV.  The numerical results of the amplitude of each graph and
the cross section at one point in the phase space are given in Table \ref{tab:testsNLG} for
(case1) $\tilde c_{12}^{\tau\tau}$=0 and (case2) $\tilde c_{12}^{\tau\tau}$=1000.  The total values
in two cases agree up to 31 digits, so the NLG invariance of this process is confirmed in tree-level order.

\begin{table}[htb]
\centerline{\begin{tabular}{lcl}
\hline
& Graph  & Absolute value of the amplitude \\\hline
\rule{0mm}{11pt}
case1: & 1 & 1.6624897226795816149294531250854308$\times 10^{-3}$ \\
& 2 & 8.1590691554170053568905511607157850$\times 10^{-15}$ \\
& 3 & 1.9404461824036032871287809608655740$\times 10^{-10}$ \\
& 4 & 2.0390333192991456823825251175977558$\times 10^{-9}$ \\
& 5 & 5.6886720681154757343568160836757535$\times 10^{-14}$ \\ \cline{2-3}
\rule{0mm}{11pt}
& total & 1.6624919797058214816061810564292905$\times 10^{-3}$ \\
\hline
\rule{0mm}{11pt}
case2: & 1 & 1.6625481763453484261164965199981440$\times 10^{-3}$ \\
& 2 & 8.1590691554170053568905511607157850$\times 10^{-15}$ \\
& 3 & 1.9404461824036032871287809608655740$\times 10^{-10}$ \\
& 4 & 2.0390333192991456823825251175977558$\times 10^{-9}$ \\
& 5 & 5.8453722653531868198152256247220933$\times 10^{-8}$ \\ \cline{2-3}
\rule{0mm}{11pt}
& total & 1.6624919797058214816061810564293247$\times 10^{-3}$ \\
\hline
\end{tabular}}
\caption{Tests for the NLG invariance}
\label{tab:testsNLG}
\end{table}

\section{Summary}
We have developed the program package {\tt GRACE/SUSY-loop} for the EW corrections
and QCD corrections of the MSSM amplitudes in one-loop order. Then we have calculated
the radiative corrections to production processes and decay processes of SUSY particles
in the framework of the MSSM using {\tt GRACE/SUSY-loop}.  We have also developed
a version of {\tt GRACE/SUSY-loop} for the extended NLG formalism by including bilinear
forms of sfermions with new NLG parameters, and have carried out tests for it in tree-level order.

\section*{Acknowledgments}

This work is partially supported by Grant-in-Aid for Scientific Research(B) (20340063) and
Grant-in-Aid for Scientific Research on Innovative Areas (21105513).



\begin{footnotesize}


\end{footnotesize}


\end{document}